# Interference-mediated intensity modulation of spin waves

Sankha Subhra Mukherjee,[1] Jae Hyun Kwon,[1] Mahdi Jamali,[1] Masamitsu Hayashi,[2] and Hyunsoo Yang[1,*]

[1]Department of Electrical and Computer Engineering and NUSNNI-NanoCore, National University of Singapore, 117576, Singapore

[2] National Institute for Materials Science, Tsukuba 305–0047, Japan

The modulation of propagating spin-wave amplitude in $Ni_{81}Fe_{19}$ (Py) films, resulting from constructive and destructive interference of spin wave, has been demonstrated. Spin waves were excited and detected inductively using pulse inductive time domain measurements. Two electrical impulses were used for launching two interfering Gaussian spin wave packets in Py films. The applied bias magnetic field or the separation between two pulses was used for tuning the amplitude of the resulting spin wave packets. This may thus be useful for spin wave based low-power information transfer and processing.

## I. INTRODUCTION

Spin waves have been identified as promising candidates for information transfer[1, 2], quantum[3] and classical[4-8] information processing, control of THz dynamics[9], and phase-matching of spin-torque oscillators[10]. Spin waves form the basis for spin-pumping[11], and have been used for the explanation[12] of the spin Seebeck effect[13]. Information transfer via spin waves does not suffer from phonon mediated joule heating in the same way as does charge transfer in the diffusive transport regime. However, in metallic systems such as permalloy, attenuation of spin waves is significant, and hence for applications involving spin waves in such systems, a method of spin wave amplification would be greatly beneficial. Previously, nonlinear parametric pumping[14] has been used for amplifying spin wave signals. However, in these methods, significant care has to be taken to make sure that the frequency of the pumping signal is precisely twice that of the signal that needs to be amplified. This allows only a single frequency to be amplified at any time. Also, the circuitry involved in the amplification process, such as an open dielectric resonator may become prohibitively complicated for most applications. Furthermore, since the amplification process is inherently nonlinear, extra spurious frequencies are produced, which might adversely affect the usefulness of the spin waves in various applications, such as in spin wave circuits. Recently, amplification has also been achieved by thermal-spin transfer torque in YIG.[15] Although this is a significant scientific demonstration, it is still not the most practical method of achieving amplification.

In this paper, a method of spin wave amplitude modulation is presented by the linear superposition of spin waves. Resonant excitation of spin dynamics has previously been exploited for reducing the power requirements of current driven domain wall motion by Thomas *et al.*,[16] and for spin transfer torque (STT) induced switching of magnetic tunnel junction (MTJ) devices by Garzon *et al.*.[17] We use interfering spin waves resulting from two closely spaced voltage impulses for the modulation of the magnitude of the resultant spin wave packets. Although spin wave interference has been studied in theory[18], simulations[19, 20], demonstrated in optical measurements[21, 22] and generation of phase shift keying signals[23] before, there is little study about quantitative time-domain electrical measurements of spin wave interference. We demonstrate how the applied bias magnetic field or the interval between two adjacent pulses can be effectively used for the amplification and attenuation of spin wave signals.

## II. EXPERIMENTAL DETAILS

Figure 1(a) shows the optical micrograph of the device used for studying spin wave amplitude modulation. A 150 μm × 40 μm × 20 nm $Ni_{81}Fe_{19}$ (Py) strip was patterned on a $Si/SiO_2$ (100 nm) substrate. A 30 nm $SiO_2$ layer was sputter deposited on top of the Py layer, and subsequently, Ta (5 nm)/Au (85 nm) was sputter deposited, and patterned into asymmetric coplanar strips (ACPS). The distance between the source lines of the excitation and detection ACPS is 10 μm. The width of the signal and ground arms of the ACPS is 10 μm and 30 μm, respectively, and the distance between the two is 5 μm. Voltage pulses applied at one of the waveguides launch a Gaussian spin wave packet[24, 25], and may be inductively detected by the other waveguide. Voltage pulses were applied by an Agilent 81134A pulse generator, and a Tektronix DPO 70604B real-time oscilloscope was used for measuring the inductive voltage generated at the detection waveguide. A 20 dB low noise amplifier was used for the amplification of the output signal. The output signals were averaged 10000 times to improve the signal-to-noise ratio. During the measurement, an out-of-plane bias magnetic field ($H_b$) was applied. The signal obtained for no applied bias field is used as the background signal, and is subtracted from the signals obtained at all other bias fields in order to obtain clean spin wave packets at each bias field. The frequencies of the resultant signals were calculated by fast Fourier transform (FFT) of the measured time domain signal. To confirm that the measured signals were indeed spin waves, the relationship between the frequency of the measured signals resulting from a single pulse excitation and the applied bias field is shown in Fig. 1(b), and shows a distinct change with applied bias field as has been shown in other reports on spin waves[24, 26]. The dependence is approximated by a second degree polynomial shown by the red solid line in Fig. 1 (b). This frequency dependence of the spin waves with the bias magnetic field is used for all subsequent calculations. For the study of the interference, Gaussian spin wave packets generated from one and two pulse excitations have been studied.

### III. SPIN WAVE INTERFERENCE MODEL

The precession frequency ($\omega$) which has been previously fitted with a second degree polynomial as mentioned above, wave vector ($k$), and the group velocity $v_g$ ($d\omega/dk$) of the spin wave packets are a function of $H_b$. A Gaussian spin wave packet may be written as $f_G(t) = A \exp\left[-(t-t_p)^2/2\sigma^2\right]\cos(kx-\omega t+\phi)$, where $A$ is the field- and position-dependent amplitude of the Gaussian wave packet, $t_p$ is the temporal position of the peak of the Gaussian wave packet, $\sigma$ is the field- and position-dependent standard deviation of the Gaussian wave packet, and $\phi$ is the phase of the sinusoidal signal. The phase $\phi$ is assumed to be constant in wave packets generated at different times. A signal excited at $t_1$ may be written as $f_G(t-t_1)$. Spin wave packets originating temporally close to one another interfere linearly when the applied excitation is in the linear regime as $f_{Tot}(t) = f_G(t) + f_G(t-t_1)$. When the phases of the sinusoidal components in the neighboring spin wave packets match, the waves constructively interfere, while, when they are out-of-phase, they destructively interfere. Thus, conditions of constructive and destructive interference may be obtained by finding phase relationships between the sinusoidal parts of the wavefunction alone, temporarily neglecting the nonlinear Gaussian dependencies. From simple trigonometric relations, it is possible to obtain the resultant interference proportionality constant as, $f_{Tot}(t) \propto |2\cos(0.5\omega t_1)|$. The relationship between the bias field and the frequency has been already measured and fitted in Fig. 1(b). Thus, for the proper separation between two consecutive input pulses, one should be able to obtain both constructive as well as destructive interference over a range of applied bias field. In order to obtain destructive interference at 2.5 GHz for example, $t_1$ should have a value of ($2\pi f t_1 = \pi$) 200 ps. However, due to the Gaussian envelopes, it is difficult to obtain an analytical expression for interference and hence a numerical solution has been sought. In this work, numerical solutions have been compared with the measured data.

## IV. EXPERIMENTAL RESULTS AND ANALYSIS

### A. Experimental data at different bias fields

A single-pulse excitation has a pulse width ($t_0$) of 100 ps (in the pulse mode) and a voltage of 2 V. A double-pulse excitation is two single-pulse excitations separated by a time period ($t_1$) of 200 ps, created by combining two 100 ps signals from two channels with a combiner. These are shown at the center of Fig. 1. Measured spin waves resulting from the single-pulse excitation at a bias field of -2.46 kOe is shown in Fig. 1(c) in red, and a simulated Gaussian wave packet is shown in blue. When two wave packets generated 200 ps apart interfere at that particular field, they destructively interfere. The measured value of this interference is shown in Fig. 1(d) by a green solid line. The result of a simulated interference between Gaussian packets 200 ps apart is shown by a blue solid line in Fig. 1(d). There is significant similarity between the simulated and the measured signals. The simulated signal comprises of two small envelopes and is zero at the center (marked by $t_c$). This is the point at which the magnitudes of the Gaussian wave packets exactly cancel each other, and thus becomes zero. This point is 100 ps from the center of either of the Gaussian wave packets, leading the center of one of the wave-packets, and trailing the other. At $t > t_c$, the spin wave packet launched at a later time has a larger amplitude than that launched earlier. Hence, the characteristics of the interference pattern beyond $t_c$ are that of the spin wave packet launched later. Similarly, at $t < t_c$, the spin wave packet launched at an earlier time has the larger amplitude, and hence, the characteristics of the interference pattern before $t_c$ corresponds to that of the earlier spin wave. During destructive interference, the spin wave packets are out-of-phase by $\pi$ with respect to one another. Therefore, as the characteristics of the resultant spin wave packet after interference changes characteristics from one wave-packet before $t_c$ to another after $t_c$, there is an abrupt phase change of $\pi$ in the resultant wave packet at $t_c$. The measured signal, shown by the green solid line in Fig. 1(d) is characteristically similar to the simulated signal. The two envelopes are separated at the center by complete destructive interference by a $\pi$ phase shift, which is a clear indication of destructive interference.

At -3.5 kOe, the resulting spin wave signal from a single-pulse excitation is shown by a red solid line in Fig. 1(e), along with a simulated result for the same field shown by a solid blue line. At this field, the wave packets originating from the double-pulse excitation constructively interfere, and the result is shown in Fig. 1(f). A blue solid line shows the simulated result, while a green solid line shows the measured data. For constructive interference, the wave-packets are in-phase, and as a result, there is no abrupt phase change in the resultant signal. Furthermore, the total amplitude of the resultant interference is greater than that resulting from a single pulse.

Unfortunately, simple Gaussian wave packets cannot be used for obtaining very accurate descriptions of the interference, especially in the low bias field regions. This is because the Gaussian pulses are created with rectangular pulses, and are actually composed of two Gaussian wave packets, one resulting from the rising edge of the pulse, and another from the falling edge of the pulse, and hence the initial rise of the Gaussian wave packets is more abrupt than the trailing edges[27]. A better description of the wave amplitudes at low fields may be obtained by taking the frequency transform of the two pulses directly. This gives additional insights into the method in which constructive and destructive interference intensities may be calculated. It is known that the Fourier transform of a rectangular pulse with a pulse width $t_0$ is $y_1 = \sin(\omega t_0/2)/(\omega t_0/2)$. It is also known that the Fourier transform of two pulses separated from one another by $t_1$ is $y_3 = y_1 \times y_2$, where $y_2 = [1+\exp(-j\omega t_1)]$. Calculated values of $|y_1|$, $|y_2|$, and $|y_3|$ are plotted as a function of frequency in Fig. 2, for $t_0 = 100$ ps, and $t_1 = 200$ ps. It is worth noting that the frequency characteristics of $y_1$ depend upon $t_0$ alone and that of $y_2$ depend upon $t_1$ alone. Hence, effective independent control of both attenuation and amplification frequencies may be obtained.

The contour plot of measured spin wave packets originating due to a single 100 ps pulse is shown in Fig. 3(a). The frequency of the measured signals increases with the magnetic field, and temporal widths between two subsequent peaks become smaller. In Fig. 3(b), the FFT of the time-domain signal shown in Fig. 3(a) is plotted. Spin waves originating from two 100 ps voltage pulses separated from each other by 100 ps (i.e. $t_1 = 200$ ps) are plotted in Fig. 3(c). The FFT of the time-domain signals resulting from two pulses is shown in Fig. 3(d). At bias fields above 3 kOe, one is clearly able to see an enhancement in the signal levels in comparison with spin wave signals arising due to the single pulse excitation.

**B. Numerical analysis**

The magnitude $m_i(H_b)$ of the signal level originating from one- ($i=1$) and two-pulse ($i=2$) excitations at a particular magnetic field, is calculated as the difference between the maximum and the minimum value of the measured signal at that magnetic field. This is plotted as a function of the bias field, for signals obtained for the single- and double-pulse excitations in Fig. 4(a). Note that the magnitude of the measured signal is dependent upon the excitation efficiency of the particular waveguides that have been used, resulting in the change in the intensity of $m_i(H_b)$, as shown. For bias fields less than 3 kOe, the magnitude of the spin wave packets due to the single-pulse is greater than that of double-pulses. However, for bias fields between 3 kOe and 4.6 kOe, the magnitude of the spin wave packets due to double-pulses constructively interfere and the resultant magnitude become greater than that due to a single-pulse. For comparing the effect of the interference in the spin wave amplitude, the magnitude of the Gaussian wave packets originating from double-pulse excitations is normalized by those originating from single-pulse excitations as $[m_2(H_b)/m_1(H_b)]$, and is plotted in Fig. 4(b) as open squares. In the same figure, the result obtained from numerical analysis is also plotted as a thick red solid line, and shows a reasonably good agreement with experiment. Figure 4 shows that the magnitude of the signal due to interference changes regularly over the magnetic field. The largest increase in signal amplitude predicted by simulation is two-fold. The value of $|y_3|$ is also plotted as a function of the applied field as a thin blue solid line, and as discussed previously, better estimating the value of the interference at smaller bias fields. Measured signals are slightly larger than simulated signals at large values of bias fields, probably due to nonlinear mixing. This allows for the field dependent control of the magnitude of the spin wave signal, and hence can be used as a spin wave modulator.

**C. Experimental data at different pulse separation**

It is also important to note that the concept of the electrical modulation of spin waves using two subsequent pulses is general and can be applied, when the external bias field is applied in other direction. For example, it is possible to apply an in-plane bias field along the signal line and as a result obtain the surface mode spin wave transport at much lower fields. Neither is the bias field the sole parameter responsible for the generation of interference. The separation between two pulses is also a very effective way for tuning the modulation resulting from the interference.

For demonstrating this phenomenon, a fixed in-plane bias field of 41 Oe is applied along the direction of the signal line and the separation between two pulses is varied from -5 to 5 ns in steps of 20 ps. Both 'unipolar' and 'bipolar' pulses are used. Unipolar pulses comprise of two consecutive pulses having the same polarity, while bipolar pulses comprise of two consecutive pulses having opposite polarity. Schematic representations of both unipolar and bipolar pulses are shown in Fig. 5. Two 100 ps bipolar pulses separated from one another by 5 ns are applied to one of the ACPS, and results in the generation of two Gaussian wave packets 5 ns apart, as shown in the lowest line plot in Fig. 5(a). When the pulses are separated from each other by 0 ns, they destructively interfere, while, when they are separated by 200 ps, they constructively interfere. The contour plot of all measurements is shown in Fig. 5(b) in the case of bipolar pulses. The interference of the two wave packets is clearly visible at the center of the contour plot. Measurement data corresponding to unipolar pulses are shown in Fig. 5(c) and (d) respectively. As can be seen, there is $\pi$ phase shift in the interference output signals for unipolar pulses compared to the bipolar case. For example, they destructively (constructively) interfere for unipolar (bipolar) pulses for a $t_\delta$ of ±200 ps. To quantify the modulation, the magnitude of the peak to peak amplitude $V_{Amp}$ is plotted as a function of $t_\delta$ in Fig. 5(e) for bipolar (unipolar) pulses in red (blue). As can be seen, the nominal amplitude of 8.97 mV, corresponding to a non-interacting wave packet, changes between 17.73 mV and 0.63 mV due to constructive and destructive interference, respectively.

**D. Micromagnetic simulations**

To better understand this behavior, we have performed micromagnetic simulations. The structure that we have used in our simulations is 6 µm in length, 4.4 µm in width, and 20 nm in thickness to preserve the aspect ratio of length over width of the actual sample. The simulation cell size is 10×10×20 nm$^3$ and is made of Permalloy (Py), having a saturation magnetization ($M_S$) of 860×10$^3$ A/m, an exchange stiffness ($A_{ex}$) of 1.3×10$^{-11}$ J/m, and a Gilbert damping constant ($\alpha$) of 0.01. We have used the object oriented micromagnetic framework (OOMMF) code for simulations that solves the *Landau-Lifshitz-Gilbert* (LLG) equation[28, 29]. In order to generate spin waves, a pulse magnetic field with rise and fall times of 60 ps and a pulse width of 80 ps was applied to a 20×4400×20 nm$^3$ volume at the center of the Permalloy film, and the spin waves were measured 1.5 μm away from the excitation source. A bias magnetic field of 200 Oe was applied to the sample along the Permalloy width during the simulations.

We have performed the simulations for two pulses with opposite voltage polarities with different time intervals between the two pulses. As can be seen in Fig. 6(a), for ±100 ps time interval between the two pulses, we have constructive interference between the two spin wave packets generated by the two pulses, while for ±200 ps one can observe destructive interference between these spin wave packets. When $2\pi t_\delta \cdot f = (2n)\pi$, where $t_\delta$ is the time interval, $f$ is the spin wave frequency, and $n$ is an integer number, a destructive interference pattern results from the two spin wave packets. When $2\pi t_\delta \cdot f = (2n+1)\pi$, a constructive interference is observed between the spin wave packets. In Fig. 6(b), we have simulated the spin wave profile for various time intervals between the two pulses at a constant bias field of 200 Oe. Clear constructive and destructive interference patterns are observable depending upon the phase difference between the two spin wave packets. Furthermore, for well separated pulses ($\Delta t > 1$ ns), the two spin wave packets propagate independently from one another. We have also performed the simulations for two pulses with the same voltage polarities as shown in Fig. 6(c) and 6(d). In contrast to the case for pulses with opposite voltage polarities, constructive interference is observed for $2\pi t_\delta \cdot f = (2n)\pi$, while one can observe a destructive interference pattern when $2\pi t_\delta \cdot f = (2n+1)\pi$, and is consistent with the measurement results.

## V. CONCLUSION

The spin wave amplitude modulation either by controlling the bias field or the separation of two pulses has been electrically demonstrated using spin wave interference. Constructive and destructive interference of spin wave has been utilized in Py films by the linear superposition of two spin waves. Both numerical calculation and micromagnetic simulations show good agreement with the experimental data. The concept of the electrical modulation of spin waves using two subsequent pulses is general and can be applied to various spin wave modes. This work lays the foundation for energy efficient information transfer as well as information processing in magnonic systems.

**ACKNOWLEDGMENTS**

S.S.M. and J.H.K. contributed equally to this work. This work is supported by the Singapore National Research Foundation under CRP Award No. NRF-CRP 4-2008-06.

[*] Electronic address: eleyang@nus.edu.sg

Fig. 1. (a) An optical micrograph of the device used for the inductive measurements of spin waves comprising of a Py strip and ACPS patterned on top of it. (b) The FFT of resultant spin-waves as a function of applied bias field. Schematic representations of input excitations have been shown below (a) and (b). These pulses are not to scale. (c) Measured (red line) and simulated (blue line) signals for spin wave packets resulting from a single-pulse excitation at -2.46 kOe. (d) Measured (green line) and simulated (blue line) signals for spin wave packets resulting from a double-pulse excitation at -2.46 kOe showing destructive interference. (e) Measured (red line) and simulated (blue line) signals for spin wave packets resulting from a single-pulse excitation at -3.5 kOe. (f) Measured (green line) and simulated (blue line) signals for spin wave packets resulting from a double-pulse excitation at -3.5 kOe showing constructive interference.

Fig. 2. Calculated values of the frequency components of two similar pulses separated in time. The frequency components of a single pulse $y_1$ (green dash dotted line), the frequency components of the two impulses separated one from the other $y_2$ (blue dashed line), and the product of the two, yielding the frequency components of two rectangular pulses separated from one another $y_3$ (red solid line).

Fig. 3. Contour plots (a) of the spin wave signal and the FFT (b) of the time-domain signal due to a single-pulse excitation. Contour plots (c) of the spin wave signal and the FFT (d) of the time-domain signal due to a double-pulse excitation. The scale bar is in mV.

Fig. 4. (a) The magnitude of a spin wave signal measured as the difference between the maximum and the minimum value of the signal at a particular bias field is plotted, using open

squares for the single-pulse excitation, and open circles for double-pulse excitations. (b) $m_2/m_1$ (open squares), $y_3$ (thin blue solid line), and $f_{Tot}(H_b)$ (thick red solid line) are plotted as a function of bias field.

Fig. 5. (a) The time-domain spin wave voltages measured for two bipolar input pulses separated from each other ($t_\delta$) by -200 ps, 0 ns, and -5 ns. A schematic representation of unipolar and bipolar input pulses is shown. (b) The contour plot of measured spin wave signals as a function of $t_\delta$ clearly shows an interference pattern resulting from bipolar pulses. (c) and (d) represent the same measurements depicted on (a) and (b) respectively, resulting from unipolar pulses. (e) The amplitude of the measured spin wave signals is seen to be modulated from its value of 8.97 mV due to interference for both unipolar (blue), and bipolar (red) pulses.

Fig. 6. (a) Simulated spin waves resulting from two bipolar pulses applied at a time-difference ($t_\delta$) of 200, 100, 0, -100, and -200 ps from one another, showing constructive and destructive interference patterns. (b) A contour plot of the interference of spin wave packets from two bipolar pulses as one of the inputs is shifted from the other by $t_\delta$. (c) Simulated spin waves resulting from two unipolar pulses applied at $t_\delta$. (d) A contour plot of the interference of spin wave packets from two unipolar pulses. The color bar is in arbitrary units.

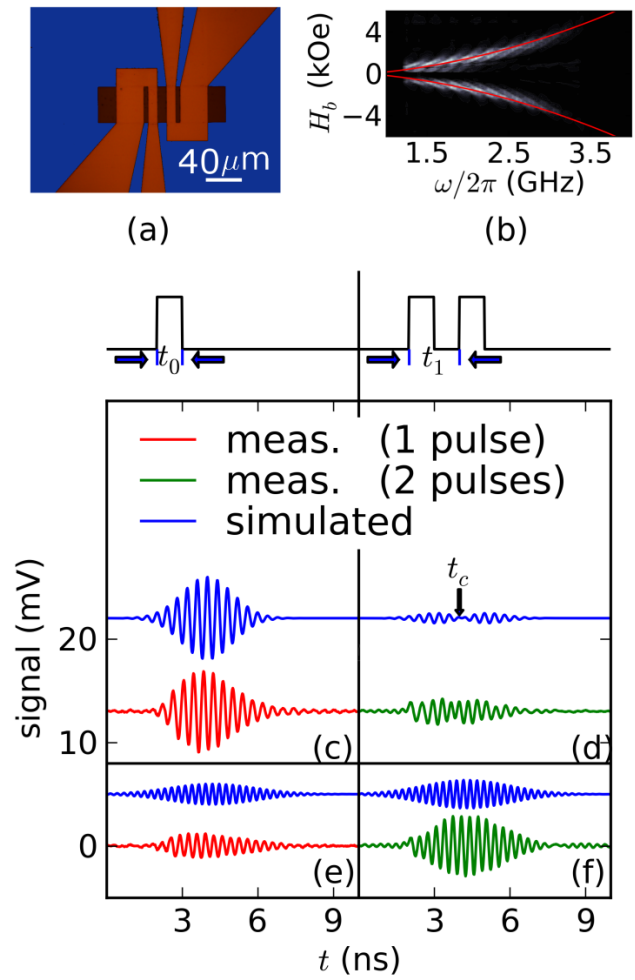

Figure 1.

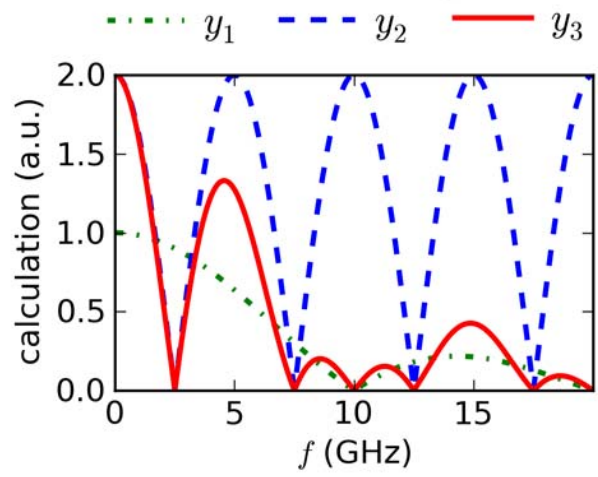

Figure 2.

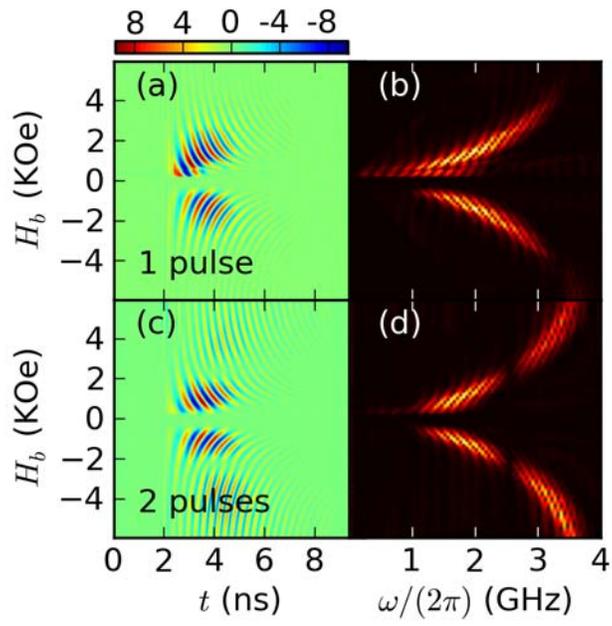

Figure 3.

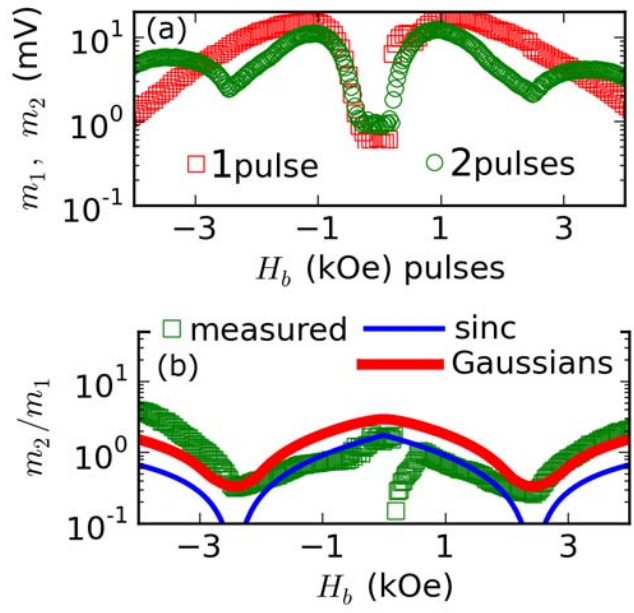

Figure 4.

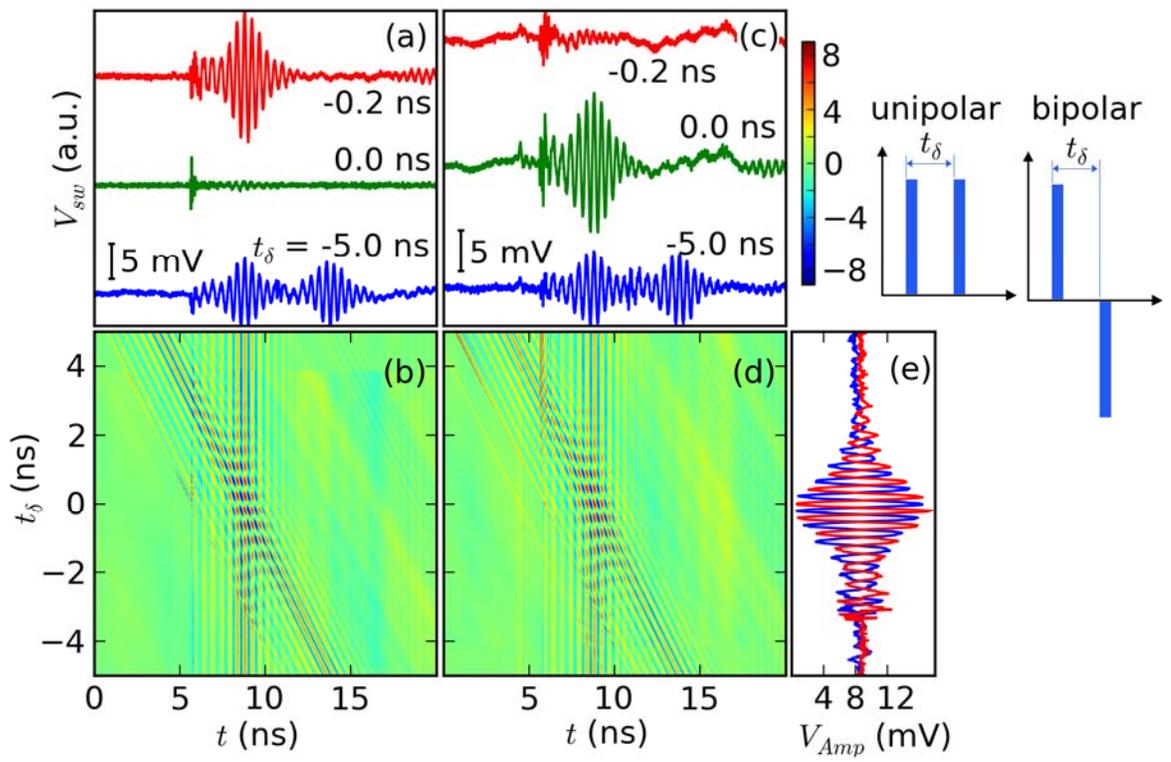

Figure 5.

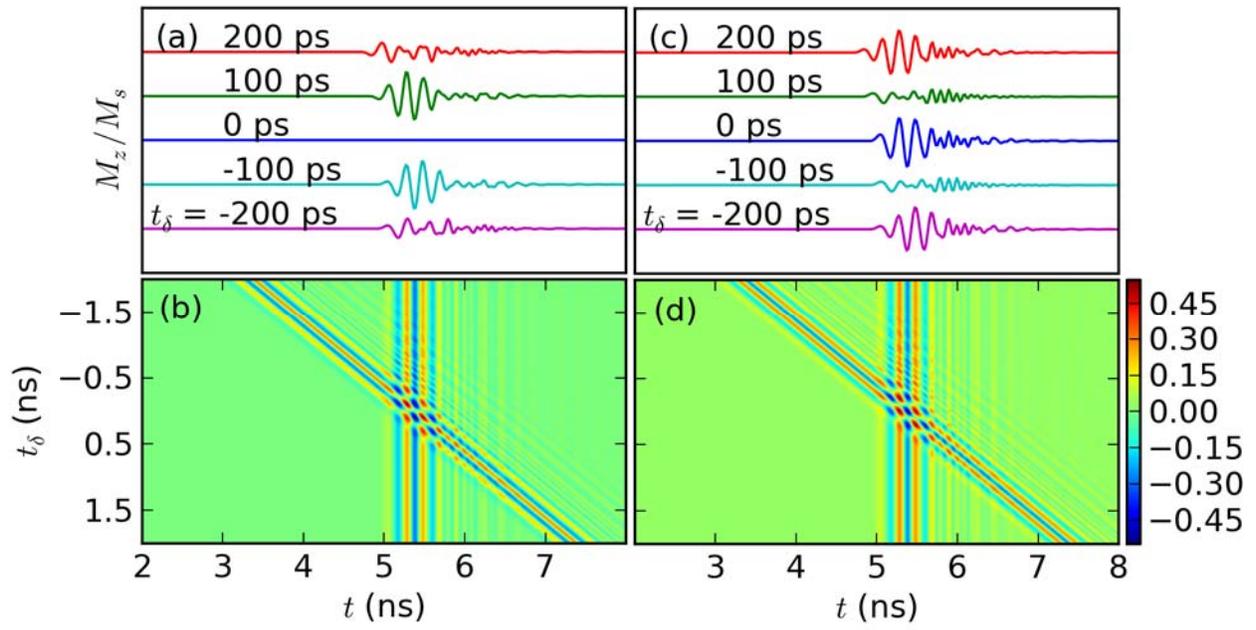

Figure 6.